\documentstyle[12pt,iopfts]{ioplppt}       

 \eqnobysec

\begin{document}

\jl{6}    

\title{Qualitative Analysis of \\
Viscous Fluid Cosmological Models \\
satisfying the Israel-Stewart \\
theory of Irreversible Thermodynamics}[Causal Viscous Fluid FRW Models]
\author {A A  Coley and R J  van den Hoogen }
\address{Department of Mathematics, Statistics, and Computing Science, Dalhousie University, \\ Halifax, Nova Scotia, Canada, B3H 3J5}
\date{\today}

\begin{abstract}
 Isotropic and spatially homogeneous viscous fluid cosmological
 models are investigated using the truncated Israel-Stewart
 [W. Israel, {\it Ann. Phys.}, {\bf 100} 1976; W. Israel and 
 J.M. Stewart, {\it Proc. R. Soc. Lond. A.}, {\bf 365} 1979; 
 ibid. {\it Ann. Phys.}, {\bf 118} 1979] theory of irreversible 
 thermodynamics to model the bulk viscous pressure.  The governing 
 system of differential equations is written in terms of dimensionless 
 variables and a set of dimensionless equations of state is then 
 utilized to complete the system.  The resulting dynamical system is
 analyzed using geometric techniques  from dynamical systems theory 
 to find the qualitative behaviour of the Friedmann-Robertson-Walker 
 models with bulk viscosity.  In these models there exists a free 
 parameter such that the qualitative behaviour of the models can be 
 quite different (for certain ranges of values of this parameter) from that
 found in models satisfying the Eckart theory studied previously.
 In addition, the conditions under which the models inflate are 
 investigated.

 \end{abstract}


 \maketitle


\section{Introduction}

Recently, spatially homogeneous imperfect fluid models have been investigated using techniques from dynamical systems theory \cite{Coley94a,Abolghasem93,Coley92,Burd94}. 
 In these papers, dimensionless variables and a set of dimensionless equations of state were employed to analyze various spatially homogeneous imperfect fluid  cosmological models.
It was also assumed that the fluid is moving orthogonal to the homogeneous spatial hypersurfaces; that is, the fluid 4-velocity, $u^a$, is equal to the unit normal of the spatial hypersurfaces.
The energy-momentum tensor can be decomposed with respect to $u^a$ according to \cite{MacCallum73}:
\begin{equation}
T_{ab}=(\rho+\bar p)u_au_b+\bar p g_{ab} + q_au_b + u_aq_b+\pi_{ab},\label{emtensor}
\end{equation}
where $\bar p \equiv p + \Pi$ and 
$\rho$ is the energy density, 
$p$ is  the thermodynamic pressure,  
$\Pi$ is the bulk viscous pressure, 
$\pi_{ab}$ is the anisotropic stress, and 
$q_a$ is the heat conduction vector as measured by an observer moving with the fluid.

In papers \cite{Coley94a,Abolghasem93,Coley92,Burd94} it was also  assumed that there is a linear relationship between the bulk viscous pressure $\Pi$ and the expansion $\theta$ (of the model),  and a linear relationship between the anisotropic stress $\pi_{ab}$ and the shear $\sigma_{ab}$; that is, 
\numparts\label{linear rel}
\begin{eqnarray}
\Pi&=&-\zeta \theta  \label{eckart}\\
\pi_{ab}&=& -2\eta\sigma_{ab} \label{viscosity approx}
\end{eqnarray}
\endnumparts
where $\zeta$
 denotes the bulk viscosity coefficient and $\eta$ denotes the shear viscosity coefficient.  
Equations (\ref{eckart}) and (\ref{viscosity approx}) describe Eckart's theory of irreversible thermodynamics. Eckart's theory is a first order approximation of the viscous pressure $\Pi$ and the anisotropic stress $\pi_{ab}$ and is assumed to be valid near equilibrium \cite{Eckart}. 
Coley and van den Hoogen \cite{Coley94a} and Coley and Dunn \cite{Coley92} have studied the Bianchi type V cosmological models using equations (1.2) as an approximation for the bulk viscous pressure and the anisotropic stress.  They found that if the models satisfied the weak energy conditions, then the models necessarily isotropize to the future.  Belinskii and Khalatnikov \cite{Belinskii76}, with different assumptions on the equations of state, found similar behaviour present in the Bianchi type I models.  The addition of viscosity allowed for a variety of different qualitative behaviours (different from that of the corresponding perfect fluid models).  
However, since the models studied in \cite{Coley94a,Abolghasem93,Coley92,Burd94} satisfy Eckart's theory of irreversible thermodynamics \cite{Eckart}, they suffer from the property  that signals in the fluid can propagate faster than the speed of light (i.e., non-causality), and also that the equilibrium states in this theory are unstable (see Hiscock and Salmonson \cite{Hiscock91} and references therein).  
Therefore,  a more complete theory  of irreversible thermodynamics is necessary for fully analyzing cosmological models with viscosity. 

 Among the first to study irreversible thermodynamics  were Israel \cite{Israel76} and Israel and Stewart \cite{Israel79,IsraelStewart79}. They included additional linear terms in the relational equations (1.2).
 Assuming that the universe can be modelled as a simple fluid and omitting certain divergence terms, the linear relational equations for the bulk viscous pressure, the heat conduction vector, and the shear viscous stress are \cite{IsraelStewart79}:
\numparts
\begin{eqnarray}
\Pi &=& -\zeta(u^a_{\:;a}+\beta_0\dot\Pi-\alpha_0q^a_{\:;a}), \label{israel1}\\
q^a &=& -\kappa T h^{ab}(T^{-1}T_{;b}+\dot u_b+\beta_1\dot q_b-\alpha_0 \Pi_{;b}-\alpha_1\pi^c_{\:b;c}),\\
\pi_{ab}&=& -2\eta\langle u_{a;b} +\beta_2\dot\pi_{ab}-\alpha_1q_{a;b}\rangle,
\end{eqnarray}\label{israel}
\endnumparts
where $h_{ab}=u_au_b+g_{ab}$ is the projection tensor and 
$\langle A_{ab}\rangle\equiv \frac{1}{2}h^c_{\:a}h^d_{\:b}(A_{cd}+A_{dc}-\frac{2}{3}h_{cd}h^{ef}A_{ef})$.
The variable $T$ represents the temperature, $\kappa$ represents the thermal conductivity, $\beta_0$, $\beta_1$ and $\beta_2$ are proportional to the relaxation times, $\alpha_0$ is a coupling parameter between the heat conduction and the bulk viscous pressure and $\alpha_1$ is a coupling parameter between the shear viscous stress and the heat conduction.
We shall refer to equations (1.3) as the truncated Israel-Stewart equations.  These   equations, (1.3), reduce to the Eckart equations (1.2) used in \cite{Coley94a,Abolghasem93,Coley92,Burd94} when $\alpha_0=\alpha_1=\beta_0=\beta_1=\beta_2=0$.

 Belinskii et al. \cite{Belinskii80} were the first to study  cosmological models satisfying the truncated Israel-Stewart theory of irreversible thermodynamics.   Using qualitative analysis, Bianchi I models were investigated with a relational equation for the bulk viscous pressure and the shear viscous stress of the form (1.3).  They also assumed equations of state of the form \cite{Belinskii80}
\begin{equation}
\zeta=\zeta_0\rho^{m}, \quad \eta=\eta_0\rho^{n}, \quad \beta_0 = \rho^{-1}, 
\quad  \text {and} \quad \beta_2 = \rho^{-1}, \label{Bel eqs state}
\end{equation}
where $m$ and $n$ are constants and $\zeta_o$ and $\eta_o$ are parameters.
The isotropizing effect found in the Eckart models no longer necessarily occurred in the truncated Israel-Stewart models.
It was also found that the cosmological singularity still exists but is of a new type, namely one with an accumulated  ``visco-elastic'' energy \cite{Belinskii80}.  Similar to the work done by Belinskii et al. \cite{Belinskii80}, Pav\'on et al. \cite{Pavon90a} and Chimento and Jakubi \cite{Chimento93} studied the flat Friedmann-Robertson-Walker (FRW) models.
  They assumed the same equations of state as Belinskii et al. \cite{Belinskii80}, namely \eref{Bel eqs state}, but studied the models using slightly different techniques.  Chimento and Jakubi \cite{Chimento93} also found exact solutions in the exceptional case $m=1/2$ (which will be of interest later). 
 They found that the future qualitative behaviour of the model was independent of the value of $m$; however, to the past,  ``bouncing solutions'' and deflationary evolutions are possible \cite{Chimento93}.

In addition, Hiscock and Salmonson \cite{Hiscock91} investigated further generalizations of the truncated Israel-Stewart theory of  irreversible thermodynamics.  
Namely, they included the  non-linear divergence terms in the relational equations (1.3).  We shall refer to such a theory as the ``Israel-Stewart-Hiscock'' theory.  
Hiscock and Salmonson used equations of state arising from the assumption that the fluid could be modelled as a Boltzmann gas.  
They concluded that when the Eckart equations, (1.2), or the truncated Israel-Stewart equations, (1.3), were used, inflation could occur, but if  the non-linear terms of the Israel-Stewart-Hiscock theory were included, inflation was no longer present.  This result led them to conclude that ``inflation is a spurious effect produced from using a truncated theory'' \cite{Hiscock91}.
However, Zakari and Jou \cite{Zakari93} also employed  equations arising from  the full Israel-Stewart-Hiscock theory of irreversible thermodynamics, but assumed equations of state of the form \eref{Bel eqs state} and found that inflation was present in all three theories (Eckart, truncated Israel-Stewart, Israel-Stewart-Hiscock).
Therefore, it appears that   the equations of state chosen  determine if the 
model will experience  bulk-viscous inflation.
Romano and Pav\'on \cite{Romano94} also analyzed Bianchi III models using both the truncated Israel-Stewart theory and the Israel-Stewart-Hiscock theory.  They only analyzed the isotropic singular points, but concluded that the qualitative behaviour  of the models in the two different theories was similar in that the anisotropy of the models dies away and the de Sitter model is a stable attractor.  

In this work we use equations of state that are dimensionless,
and hence we are generalizing the work in 
\cite{Coley94a,Abolghasem93,Coley92,Burd94} in which viscous fluid cosmological models satisfying equations (1.2) were studied.
One reason for using dimensionless equations of state is that the equilibrium points of the system of differential equations describing spatially homogeneous models will represent self-similar cosmological models \cite{Coley94b}.
In addition, it could be argued that the use of dimensionless equations of state is   natural  in the sense that the corresponding physics is scale invariant  (see also arguments in Coley \cite{Coley90b}).

The intent of this work is to build upon the foundation laid by Belinskii et al. \cite{Belinskii80}, Pavon et al. \cite{Pavon90a} and Chimento and Jakubi \cite{Chimento93}, and investigate viscous fluid cosmological models satisfying the linear relational equations (1.3). We will use dimensionless variables and dimensionless equations of state to study the qualitative properties of   isotropic and spatially homogeneous cosmological models.  In particular, we shall  study this new ``visco-elastic'' singularity and  we shall determine whether bulk-viscous inflation is possible.  We will also determine if there is a qualitative difference between these models and the models studied by Burd and Coley \cite{Burd94} where the Eckart equation \eref{eckart} was assumed.  

In section~\ref{II} we define the models and establish the resulting dynamical system.
 In section~\ref{III}  we investigate the qualitative behaviour of the system for different values of the physical parameters.    In section~\ref{IV} we discuss and interpret our results and in section~\ref{V} we end with our conclusions. 
 For simplicity we have chosen units in which $8\pi G = c= 1$.


\section{Friedmann-Robertson-Walker Models\label{II}}

 In this paper we   assume that the spacetime is spatially homogeneous and isotropic and that the fluid is moving orthogonal to the spatial hypersurfaces.
 The energy-momentum tensor considered in this work is an imperfect fluid with a non-zero bulk viscosity (that is there is no heat conduction, $q_a=0$, and no anisotropic stress, $\pi_{ab}\equiv 0$).   
The Einstein field equations ($G_{ab}=T_{ab}$) and the energy conservation equations ($T^{ab}_{\:\:\: ;b}u_a=0$) can be written as (see Burd and Coley \cite{Burd94}):
\numparts
\begin{eqnarray}
\dot\theta &=& - {{1}\over3}\theta^2 - {{1}\over2}(\rho + 3\bar p),\label{raychaudhuri}\\
\dot\rho  &=&   - \theta(\rho +\bar p), \label{rho} \\ 
 \theta^2&=& 3\rho  -\frac{1}{3}\,^3\!R, \label{gen.equality}
 \end{eqnarray} 
\endnumparts
 where $\theta$ is the expansion and $\,^3\!R$ is the curvature of the spatial hypersurfaces.  If the curvature is negative, i.e., $\,^3\!R<0$, then the FRW model is open,   if $\,^3\!R \equiv 0$ then the model is flat, and if $\,^3\!R>0$ then the  FRW model is closed. 
 Assuming that the energy density, $\rho$, is non-negative, it is easily seen from \eref{gen.equality} that in the open and flat FRW models  the expansion is always non-negative, i.e. $\theta\geq0$, but for the closed FRW models the expansion may become negative. (Great care must be taken in this case because the dimensionless quantities that we will be using become ill-defined at $\theta =0$.)
 
  We can obtain an evolution equation for $\Pi$ by solving  \eref{israel1} for $\dot\Pi$, 
\begin{equation}
\dot\Pi=-\frac{\Pi}{\beta_0\zeta}-\frac{1}{\beta_0}\theta.\label{newtau}
\end{equation}

Now the system of equations defined by equations \eref{raychaudhuri}, \eref{rho}, and \eref{newtau} constitute a dynamical system of the form
 $\dot {\bf X} = {\bf F}({\bf X})$, where ${\bf X}=(\theta,\rho,\Pi)$.  
This system of equations is invariant under the mapping (see Coley and van den Hoogen \cite{Coley94a,Coley94b})
\begin{equation}
\begin{array}{llll}
\theta \to\lambda\theta,   & \Pi \to \lambda^2 \Pi ,   & p\to\lambda^2 p,& \  \\
\rho \to\lambda^2\rho,\qquad & \zeta \to \lambda \zeta, \qquad & \beta_0\to\lambda^{-2} \beta_0, & t\to{\lambda}^{-1}t,    
\end{array}\label{transformation}
\end{equation}
and this invariance implies that there exists a symmetry in the dynamical system \cite{Bluman}.  
 Therefore, we introduce new dimensionless variables $x$, $y$ and
a new time variable $\Omega$ as follows: 
\begin{equation}
x \equiv {{3\rho}\over {\theta^2}}, \qquad
  y\equiv {{9\Pi}\over{\theta^2}},\quad{\text {and} }\quad {{d\Omega}\over {dt}} = - {{1}\over 3}\theta,\label{new vars}
\end{equation}
 and consequently the Raychaudhuri equation, \eref{raychaudhuri}, effectively decouples from the system.

In order to complete the system of equations we need to specify
equations of state for the quantities $p$, $\beta_0$, and $\zeta$.
In principle
equations of state can be derived from kinetic theory, but in practice one must
specify phenomenological equations of state which may or may not have any
physical foundations.   Following Coley
\cite{Coley90b,Coley90a}, we introduce dimensionless equations of state of the
form 
\numparts
\label{equations of state} 
\begin{eqnarray}
{p\over {\theta^2}} &=& p_ox^{\ell}, \\
{{\zeta}\over {\theta}} &=& \zeta_ox^m,  \\
{{3}\over {\beta_0\theta^2}} &=& ax^{r_1},   \end{eqnarray}
\endnumparts
where $p_o$, $\zeta_o$, and $a$ are positive constants, and $\ell$, $m$, and ${r_1}$  are constant parameters ($x$ is the
dimensionless density parameter defined earlier).  In the models under consideration,
  $\theta$ is  positive in the open and flat FRW models, thus equations (2.6) are well defined. In the closed FRW model the expansion could become zero, in which case these equations of state break down.  However, we can utilize these equations to model the asymptotic behaviour at early times, i.e., when $\theta>0$.
The most commonly used equation of state for the pressure is the barotropic
equation of state $p = (\gamma - 1)\rho$, whence $p_o = {{1}\over 3} (\gamma -
1)$ and $\ell= 1$ (where $1 \leq \gamma\leq 2$ is necessary for local mechanical
stability and for the speed of sound in the fluid to be no greater than the speed of
light).
  
We define a new constant $b=a/\zeta_o$.  Using equations  (2.5) and (2.6), we find that equations \eref{rho} and \eref{newtau}  reduce to 
\numparts
\label{sys of equations}
\begin{eqnarray}
\frac{dx}{d\Omega}&=& (1-x)[(3\gamma-2)x+y], \\
\frac{dy}{d\Omega} &=& -y[2+y+(3\gamma-2)x] + bx^{{r_1}-m}y + 9ax^{r_1}.
\end{eqnarray}
\endnumparts
Also, from the Friedmann equation,  \eref{gen.equality}, we obtain 
\begin{equation}
1-x=-\frac{3}{2}\,^3\!R\theta^{-2}.\label{friedmann}
\end{equation}
Thus,  the line $x=1$ divides the phase space into three invariant sets, $x<1$, $x=1$, and $x>1$.
 If $x=1$, then the model is necessarily a flat FRW model, if $x<1$ then the model is necessarily an open FRW model, and if $x>1$ the model is necessarily a closed FRW model.

The equilibrium points of the above system all represent self-similar cosmological models, except in the case $\gamma=3\zeta_0$.  If $\gamma \not = 3\zeta_0$, the behaviour of the equations of state, equation (2.6), at the equilibrium points, is independent of the parameters $m$ and ${r_1}$; namely  the behaviour is
\begin{equation}
\zeta \propto \rho^{\frac{1}{2}}, \quad  \text{ and } \quad \beta_0 \propto \rho^{-1}.
\end{equation}
Therefore,  natural choices for $m$ and ${r_1}$   are respectively $1/2$, $1$. 
We note that in the exceptional case $\gamma=3\zeta_0$, there is a singular point $\{x=1, y=-3\gamma\}$ which represents a de Sitter solution and is not self-similar.  (This is also the case in the Eckart theory as was analyzed by Coley and van den Hoogen \cite{Coley94a}.)

To further motivate the choice  of the parameter ${r_1}$, we consider the velocity of a viscous pulse in the fluid \cite{Zakari93},
\begin{equation}
v=\left(\frac{1}{\rho\beta_0}\right)^{1/2},
\end{equation}
where $v=1$ corresponds to  the speed of light. Using \eref{new vars} and equations (2.6), we obtain
\begin{equation}
v=(ax^{{r_1}-1})^{1/2}.
\end{equation}
Now, if ${r_1}=1$ then not only do we obtain the correct asymptotic behaviour 
  of the equation of state for the quantity $\beta_0$ but we are also
allowed to   choose $a<1$ since then the velocity of a viscous pulse is less than the velocity of light for any value of the density parameter $x$.  
Thus in the remainder of this analysis we shall choose ${r_1}=1$.  In  order for the system of differential equations (2.7) to remain continuous everywhere, we   also assume $m\leq1$.


\section{Qualitative Analysis\label{III}}

\subsection{$m={r_1}=1$\label{III.1}}

We now study the specific case when $m={r_1}=1$.  In this case there are three singular points,
\begin{equation}
(0,0), \qquad (1,y^-), \quad \text{ and } \quad (1,y^+),
\end{equation}
where
\begin{equation}
\fl y^-=\frac{b-3\gamma}{2}-\frac{\sqrt{(b-3\gamma)^2 + 36 a}}{2}
\quad \text{ and } \quad
y^+= \frac{b-3\gamma}{2}+\frac{\sqrt{(b-3\gamma)^2 + 36 a}}{2}. 
\label{singular points 2} 
\end{equation}

The point $(0,0)$ has eigenvalues
\begin{equation}
\fl \quad \frac{3\gamma-4+b}{2}-\frac{1}{2}{\sqrt{(b-3\gamma)^2+36a}},\quad \frac{3\gamma-4+b}{2}+\frac{1}{2}{\sqrt{(b-3\gamma)^2+36a}}.
\end{equation}
This point is either a saddle or a source depending on the value of the  parameter, $B_1$;
\begin{equation}
B_1=(2-b)(3\gamma-2)+9a.
\end{equation}
If $B_1>0$, then the point is a saddle point and if $B_1<0$, then the point is a source. If $B_1=0$ (the bifurcation value), then the point is degenerate (discussed later).

The point $(1,y^-)$ has eigenvalues
\begin{equation}
\sqrt{(b-3\gamma)^2+36a},\quad -\frac{3\gamma-4+b}{2}+\frac{1}{2}{\sqrt{(b-3\gamma)^2+36a}}.
\label{eig1}\end{equation}
 If $B_1<0$, then the point $(1,y^-)$ is a saddle point and if $B_1>0$, then the point $(1,y^-)$ is a source. If $B_1=0$ (the bifurcation value), then the point $(1,y^-)$ is degenerate (discussed later).

The singular point $(1,y^+)$ has eigenvalues
\begin{equation}
-\sqrt{(b-3\gamma)^2+36a},\quad -\frac{3\gamma-4+b}{2}-\frac{1}{2}{\sqrt{(b-3\gamma)^2+36a}}.
\label{eig2}\end{equation}
This singular point is a sink for $\gamma > 2/3$  (See also \Tref{table I} for details).

In addition to the invariant set $x=1$, there exist two other invariant sets.
These are straight lines, $y=m_{\pm}x$,  where
\begin{equation}
m_\pm = \frac{ (b-3\gamma) \pm \sqrt{(b-3\gamma)^2+36a}}{2}.
\end{equation}
The invariant line $y=m_+x$ passes through the singular points $(0,0)$ and $(1,y^+)$ while the line $y=m_-x$ passes through the singular points $(0,0)$, and $(1,y^-)$.
These invariant sets represent the eigendirections at each of the singular points [see also \ref{appendix 1}].

In order to sketch a complete phase portrait,   we   also need to calculate the vertical isoclines which occur whenever $dx/d\Omega=0$.  From (2.7) we can see that this occurs either when $x=1$ or when $y=-(3\gamma-2)x$.  This straight line  passes through the origin, and   through the singular point $(1,y^-)$ if    $B_1=0$.  If $B_1>0$ then the vertical isocline has a negative slope which is greater than the slope of the slope of the invariant line $y=m_- x$, [i.e., $m_-<-(3\gamma-2)$], and when $B_1<0$ the vertical isocline has a negative slope which is less than the slope of the invariant line  $y=m_- x$ [i.e., $m_->-(3\gamma-2)$].

To complete the analysis of this model we need to analyze the points at infinity.
We do this by first converting to polar coordinates and then compactifying the radial coordinate.  We change to polar coordinates via
\begin{equation}
r^2=x^2+y^2 \qquad \text{and} \qquad \theta = \tan^{-1}\frac{y}{x},
\end{equation}
and we derive evolution equations for $r$ and $\theta$. We essentially compactify the phase space by changing our radial coordinate $r$ and our time $\Omega$ as follows, 
\begin{equation}
\bar r=\frac{r}{1+r} \qquad \text{and} \qquad \frac{d\Omega}{d\tau} = (1-\bar r),
\end{equation}
 that is, the plane ${\Bbb R}^2$ is mapped to the interior of the unit circle, with the boundary of this circle representing points at infinity of ${\Bbb R}^2$.  We have (for ${r_1}=1$ and general $m$)
\begin{eqnarray}
\fl \frac{d\bar r}{d\tau} = (1-\bar r)\biggl\{
\bar r(1-\bar r)[(3\gamma-2)\cos^2\theta -2\sin^2\theta+(1+9a)\cos\theta\sin\theta]\nonumber\\
  -{\bar r}^2[(3\gamma-2)\cos\theta
+\sin\theta] +{\bar r}^{2-m}(1-\bar r)^{m}[b\sin^2\theta\cos^{1-m}\theta]\biggr\}, \\
\fl \frac{d\theta}{d\tau}= (1-\bar r)\biggl\{9a\cos^2\theta-\sin^2\theta-3\gamma\cos\theta\sin\theta
+b {\bar r}^{1-m}(1-\bar r)^{m-1} \sin\theta\cos^{2-m}\theta \biggr\}.
\end{eqnarray}
We easily conclude that if $m=1$ (or any $m>0$),  then the entire circle, $\bar r =1$, is singular.  Therefore, we have  a non-isolated set of singular points at infinity.
To determine their stability we look at the sign of $d\bar r/d\tau$ as $\bar r \to 1$.  In this case we see
\begin{equation}
\frac{d\bar r}{d\tau}\Biggl|_{\bar r\approx 1}\Biggr. \approx (3\gamma-2)\cos\theta+\sin\theta
\end{equation}
which implies that points above the line $y=-(3\gamma-2)x$ are repellors, while those points which lie below the line are attractors.

For completeness, we would also like to determine the qualitative behaviour of the system at the bifurcation value $B_1 = 0$ where the singular points are $(1,b-2)$ and the line of singular points $y=-(3\gamma-2)x$. (Note that since $B_1= 0$, $b-2>0$.)  Fortunately we are able to completely integrate the equations in this case to find
\begin{equation}
\vert b-2-y \vert = k\vert 1 - x \vert,
\end{equation}
where $k$ is an integration constant.  We see that all trajectories are straight lines that pass through the point $(1,b-2)$.  It is straightforward to see that the line of singular points are repellors while the point $(1,b-2)$ is an attractor.  We are now able to  sketch  complete phase portraits  (See Figures \ref{Figure 1}, \ref{Figure 2} and \ref{Figure 3}).

\subsection{$m=1/2$ and ${r_1}=1$ \label{III.2}}

  This is a case of particular interest since it represents the asymptotic behaviour of the FRW models for any $m$ and ${r_1}$ (since at the singular points the viscosity coefficient behaves like $\zeta \propto \rho^{1/2}$ and the relaxation time like $\beta_0 \propto \rho^{-1}$).
Note that the physical phase space is defined for $x\geq0$, but the system is  not differentiable at $x=0$.
 In this case there are four singular points,
\begin{equation}
(0,0), \qquad (\bar x, \bar y),\qquad (1,y^-), \quad \text{ and } \quad (1,y^+),
\end{equation}
where
\begin{equation}
\fl \bar x = \frac{\bigl(9a+2(3\gamma-2)\big)^2}{b^2(3\gamma-2)^2}, \qquad \bar y= -(3\gamma-2)\bar x,
\end{equation}
and $y^+$ and $y^-$ are given be equation \eref{singular points 2}.

The dynamical system, (2.7) is not differentiable at the singular point $(0,0)$.  We can circumvent this problem by changing variables to
$u^2=x$ and a new time variable $\tau$ defined by $d\Omega/d\tau=u$.  The system then becomes
\numparts
\begin{eqnarray}
\frac{du}{d\tau}&=&(1-u^2)[(3\gamma-2)u^2+y], \\
\frac{dy}{d\tau}&=&u[9au^2-2y-y^2-(3\gamma-2)u^2+byu].
\end{eqnarray}
\endnumparts
In terms of the new variables the system is differentiable at the point $u=0,y=0$, but one of the eigenvalues is zero and hence the point is
not hyperbolic.  Therefore in order to determine the stability of the point we change to polar coordinates, and find   that the point has some saddle-like properties; however, the true determination of the stability is difficult.  [We investigate the nature of this singular point numerically --- see  \ref{appendix 2}.]
 
The singular point $(\bar x,\bar y)$
has eigenvalues
\begin{equation}
\fl \quad \frac{(3\gamma-2)^2+9a}{2(3\gamma-2)}
\pm\frac{1}{2}{\sqrt{
      \left(\frac{(3\gamma-2)^2+9a}{(3\gamma-2)}\right)^2  
-2(1-\bar x)[9a+2(3\gamma-2)]}}.
\end{equation}
This singular point varies both its position in phase space with its stability depending upon whether $\bar x$ is less than, equal to or greater than one.    If $B_1>0$, then   $\bar x >1$ and the point is a saddle point.  
If $B_1<0$, then   $\bar x<1$ and the point is a source.
Finally, if $B_1=0$ (the bifurcation value), then $\bar x=1$ and the point is degenerate (discussed later).

The stability of the points $(1,y^-)$ and $(1,y^+)$ is the same as in the privious case, see equations \eref{eig1} and \eref{eig2} for their eigenvalues and the corresponding text.   (See also \Tref{table I} for details).

  The vertical isoclines  occur at $x=1$ and $y=-(3\gamma-2)x$.  This straight line is easily seen to pass through the origin and the point $(\bar x, -(3\gamma-2)\bar x)$.    If $B_1>0$, then the vertical isocline lies below the point $(1,y^-)$ and if  $B_1<0$, the vertical isocline lies above the point $(1,y^-)$.  Finally if $B_1=0$ (the bifurcation value), the vertical isocline passes through the point $(1,y^-)$. 

 From an analysis similar to that in the previous subsection, we  conclude that there is a non-isolated set of singular points at infinity.  Their qualitative behaviour is the same in this case as in the previous case; namely,
  points which lie above the line $y=-(3\gamma-2)x$ are repellors, while those points which lie below the line are attractors.

At the bifurcation value $B_1=0$, the points $(1,y^-)$ and $(\bar x, -(3\gamma-2)\bar x)$ come together; consequently these points undergo a saddle-node bifurcation as $B_1$ passes through the value $0$.  The singular point is no longer hyperbolic, but the qualitative behaviour near the singular point can be determined from the fact that we know the nature of the bifurcation.  Hence the singular point is a repelling node in one sector and a saddle in the others.
A complete phase portrait is sketched in Figures \ref{Figure 4}, \ref{Figure 5} and \ref{Figure 6}.


\begin{table}
\caption{Qualitative nature of the singular points of the dynamical system, \protect(2.7), for different values of the parameter $B_1$$^{\rm a}$ and ${r_1}=1$ (with respect to $\Omega$-time).}
 \begin{indented}
\item[]\begin{tabular}{lcccc}
\br
\  &  $(0,0)$ & $(1,y^-)$$^{\rm b}$ & $(1,y^+)$$^{\rm c}$ & $(\bar x, \bar y)$ $^{\rm d}$\\
 \mr
$m=1$ and $B_1<0$ & 
source & 
saddle &
sink &
 \\
$m=1$ and $B_1=0$ & 
source $^{\rm e}$ & 
source $^{\rm e}$ &
sink &
 \\
$m=1$ and $B_1>0$ & 
saddle & 
source &
sink &
 \\
 $m=1/2$ and $B_1<0$ & 
saddle & 
saddle &
sink &
source\\
 $m=1/2$ $B_1=0$ & 
saddle &
saddle-node $^{\rm f}$ & 
sink &
saddle-node $^{\rm f}$ \\
$m=1/2$ and $B_1>0$ & 
saddle & 
source &
sink &
 saddle \\
 \br
\end{tabular}
\item [] $^{\rm a}$  {$B_1=(6-b)(\gamma-2)+3a$.} 
\item []$^{\rm b}$ {$y^-=\left({b-3\gamma}-{\sqrt{(b-3\gamma)^2 + 36 a}}\right)/{2}
$.} 
\item [] $^{\rm c}$ {$y^+=\left({b-3\gamma}+{\sqrt{(b-3\gamma)^2 + 36 a}}\right)/{2}
$.}
\item [] $^{\rm d}$ {$\bar x = {\bigl(9a+2(3\gamma-2)\big)^2}/{b^2(3\gamma-2)^2}, \qquad \bar y= -(3\gamma-2)\bar x$,}
\item [] $^{\rm e}$ {These points are part of the non-isolated line singularity $y=-(3\gamma-2)x$.}
\item [] $^{\rm f}$ {This is the situation when the points $(1,y^-)$ and $(\bar x, \bar y)$ coalesce.}
\label{table I}
\end{indented}
\end{table}


\section{Discussion\label{IV}}

\subsection{Exact Solutions}

The exact solution of the Einstein field
equations at each of the singular points represent the asymptotic solutions (both past and future) of FRW models with a causal viscous fluid source.
The solution at each of the singular points represents a self similar
cosmological model except in one isolated case [see the singular point 
$(1,y^-)$].

At the singular point $(0,0)$ we
 have (after a re-coordinatization)
 \begin{tabbing}
 $\theta(t)=3A^+(t-t_o)^{-1}$,\qquad \= $a(t)=a_o(t-t_o)$,  \kill
 $\theta(t)=3t^{-1}$, \> $a(t)=a_ot$, \\
 $\rho(t)=0$, \> $\Pi(t)=0$,
 \end{tabbing}
 which represents the standard vaccuum Milne model.

The singular point $(1,y^+)$ represents a flat FRW model with a solution
(after a re-coordinatization)
\begin{tabbing}
$\theta(t)=3A^+(t-t_o)^{-1}$,\qquad \= $a(t)=a_o(t-t_o)$,  \kill
$\theta(t)=3A^+t^{-1}$, \> $a(t)=a_ot^{A^+}$,  \\
$\rho(t)=3(A^+)^2t^{-2}$, \> $\Pi(t) = y^+(A^+)^2t^{-2}$,
\end{tabbing}
where $A^+=2/(3\gamma+y^+)>0$.

The singular point $(1,y^-)$ represents a flat FRW model. If $\gamma \not = 3
\zeta_o$ then the solution is\begin{tabbing}
 $\theta(t)=3A^+(t-t_o)^{-1}$,\qquad \= $a(t)=a_o (t-t_o)$,  \kill
$\theta(t)=3A^-(t-t_o)^{-1}$, \> $a(t)=a_o|t-t_o|^{A^-}$,  \\
$\rho(t)=3(A^-)^2(t-t_o)^{-2}$, \> $\Pi(t) = y^+(A^-)^2(t-t_o)^{-2}$,
\end{tabbing}
where $A^-=2/(3\gamma+y^-)$. (Note that in this case we cannot simply change coordinates to remove the constants of integration.)
The sign of $A^-$ depends on the sign of $\gamma-3\zeta_o$.
If $\gamma>3\zeta_o$ then $A^->0$, and if 
 $\gamma<3\zeta_o$ then $A^-<0$.  Thus  if $A^-<0$ then $\theta$ is positive only in the interval $0\leq t \leq t_o$ and hence we can see that after a finite
time $t_o$ $\theta$,  $\rho$, and $a$ all approach infinity.
(We will see later that the WEC is violated in this case).
  If $A^->0$ then we can re-coordinatize the time $t$ so as to remove the constant of integration, $t_o$, and the absolute value signs in the solution for $a(t)$.
If $\gamma=3\zeta_o$ then $A^-=0$ and the solution is the de Sitter model with
(after a re-coordinatization)
\begin{tabbing}
$\theta(t)=3A^+(t-t_o)^{-1}$,\qquad \= $a(t)=a_o(t-t_o)$,  \kill
$\theta(t)=3H_o$, \> $a(t)=a_o\e^{H_ot}$,\\ 
$\rho(t)=3{H_o}^2$, \> $\Pi(t)={H_o}^2y^-$. 
\end{tabbing}
This  exceptional solution is the only one that is not  self-similar.  It can be noted here that this is precisely the same situation
that occurred in the Eckart models studied in \cite{Burd94} and 
\cite{Coley94a}.

The singular point $(\bar x, \bar y)$ represents  either an open, flat,
or closed model depending on the value of the parameter $B_1$.  The
solution  in all cases is (after a re-coordinatization)
\begin{tabbing}
  $\theta(t)=3A^+(t-t_o)^{-1}$,\qquad \= $a(t)=a_o(t-t_o)$,  \kill
  $\theta(t)=3t^{-1}$, \> $a(t)=a_ot$,  \\
$\rho(t)=3\bar xt^{-2}$, \> $\Pi(t)=\bar yt^{-2}$. 
\end{tabbing}

\subsection{Energy Conditions}

 The weak energy condition (WEC) states that $T_{ab}W^aW^b\geq0$ for any timelike vector $W^a$ \cite{HawkingEllis}.  In the model under investigation this inequality reduces to $\rho\geq0$ and $\rho+p+\Pi\geq0$. Assuming that $p=(\gamma-1)\rho$ from here on,   the WEC in dimensionless variables becomes 
\begin{equation}
x\geq0 \quad\text{and}\quad y\geq -3\gamma x.
\end{equation}
 
The dominant energy condition (DEC) states that for every timelike $W_a$, $T_{ab}W^aW^b\geq0$ and $T^{ab}W_a$ is non-spacelike \cite{HawkingEllis}.  
Here this inequality reduces to $\rho\geq0$ and $-\rho\leq p+\Pi \leq \rho$ which when transformed to    dimensionless variables  becomes 
\begin{equation}
\text {WEC and}\quad y\leq 3(\gamma-2) x.
\end{equation}

The strong energy condition (SEC) states that  $T_{ab}W^aW^b-\frac{1}{2}T_a^{\;a}W^bW_b\geq0$  \cite{HawkingEllis}.  
Here this inequality reduces to $\rho\geq0$ and $\rho+3 p+3\Pi \geq0$ which when transformed to    dimensionless variables    becomes 
\begin{equation}
\text {WEC and}\quad y\geq -(3\gamma-2) x.
\end{equation}

If we assume that the WEC is satisfied throughout the evolution of these models
then we find that there are five distinct situations.
If $\gamma<3\zeta_o$, then $B_1>0$ and the line $y=-3\gamma x$ intersects the
line $x=1$  at a point $y>y^-$.
If $\gamma=3\zeta_o$, then $B_1>0$ and the line $y=-3\gamma x$ intersects the
line $x=1$ at the point $y=y^-$.
If $\gamma>3\zeta_o$ then $B_1$ can be of any sign or zero, but the line $
y=-3\gamma x$ intersects the line $x=1$ at a point $y<y^-$.
If  the WEC condition is   satisfied throughout the evolution of 
these models then the
possible asymptotic behaviour of the models is greatly restricted.

\subsection{Asymptotic Behaviour}

The qualitative behaviour depends  on the values of $B_1$ and $m$.
If the parameter $m$ is different from unity then there is an additional singular point.  This property is also present in the Eckart models studied by Burd and Coley \cite{Burd94} and Coley and van den Hoogen \cite{Coley94a}.
The value $m=1$ corresponds to the case when the dynamical system, (2.7), is 
polynomial (the only other value of $m$ that exhibits this property is $m=0$).  The value  $m=1/2$ is of particular interest as it represents the asymptotic behaviour of all the viscous fluid FRW models, and also, this is the case when the equation of state for $\zeta$ is independent of $\theta$ (i.e., $\zeta \propto \rho^{1/2}$).
 The    parameter, $B_1$, plays a  role similar to  the parameter $9\zeta_o-(3\gamma-2)$ found in both Burd and Coley \cite{Burd94} and Coley and van den Hoogen \cite{Coley94a}.  The value of the parameter $B_1$ determines the
stability and global behaviour of the system.

One of the goals of this paper is to determine the generic behaviour
of the system of equations (2.7).
Using the above energy conditions, and in particular the WEC, and the phase-portraits (Figures \ref{Figure 1}--\ref{Figure 6}), we can determine the generic and exceptional behaviour of all the viscous fluid models
 satisfying the WEC.
 We are primarily interested in the generic asymptotic behaviour of the FRW model with viscosity:  If we  consider the dynamical system, (2.7), as ${\bf \dot X}={\bf F}({\bf X})$ where ${\bf X}=(x,y,a,b,\gamma)\subset{\Bbb R}^5$, $x,y$ are the variables, and $a,b,\gamma$ are the free parameters
then generic behaviour occurs in sets of non-zero measure with respect to the  set ${\Bbb R}^5$  (except for the flat models in which case the state space is a subset of ${\Bbb R}^4$). For example, the case $B_1=0$ is a set of measure zero
with respect to the set ${\Bbb R}^5$.  All behaviour is summarized in \Tref{table II}.


\begin{table}
\caption{Asymptotic behaviour of the FRW models with bulk viscosity satisfying the WEC with ${r_1}=1$.}
\footnotesize\rm
\begin{tabular}{lllll}
\br
parameters & m & models & generic behaviour & exceptional behaviour$^{\rm a}$\\
\mr
$\gamma<3\zeta_o$, $B_1>0$ & 
$m=1,1/2$&
open & 
& 
$(1,y^+) \to (0,0)$ 
 \\
& 
&
flat  & 
$(1,y^+) \to \infty$ &
 \\
  &
& 
closed  & 
$(1,y^+) \to \infty$&
$(\bar x,\bar y)\to \infty$, $(1,y^+) \to (\bar x, \bar y)$$^{\rm b}$
  \\
$\gamma\geq3\zeta_o$, $B_1>0$  &
$m=1,1/2$& 
open & 
$(1,y^+) \to (1,y^-)$  &
$(1,y^+) \to (0,0)$, $(0,0)\to (1,y^-)$
 \\
  & 
&
flat  & 
$(1,y^+) \to \infty$ &
  \\
  & 
&
  & 
$(1,y^+) \to (1,y^-)$ &
  \\
  &
& 
closed  & 
$(1,y^+) \to \infty$ & 
  $(\bar x,\bar y)\to (1,y^-)$\\
  &
& 
   & 
$(1,y^+) \to (1,y^-)$ & 
 $(\bar x,\bar y)\to\infty$, $(1,y^+) \to (\bar x,\bar y)$$^{\rm b}$
\\
$\gamma>3\zeta_o$, $B_1=0$  &
m=1 & 
open & 
$(1,y^+) \to (x_o,-(3\gamma-2)x_o)$ 
 \\
  & 
&
flat  & 
$(1,y^+) \to \infty$& 
  \\
  & 
&
  & 
$(1,y^+) \to (1,y^-)$ &
  \\
  & 
&
closed  & 
$(1,y^+) \to \infty$ &
  \\
  & 
&
  & 
$(1,y^+) \to (x_o,-(3\gamma-2)x_o)$ &
  \\
  &
m=1/2   & 
open & 
$(1,y^+) \to (1,y^-)$& 
$(1,y^+) \to (0,0)$ $(0,0)\to (1,y^-) $\\
  & 
&
flat  & 
$(1,y^+) \to \infty$ &
  \\
  &
&
 & 
$(1,y^+) \to (1,y^-)$ &  
   \\
  & 
&
closed  & 
$(1,y^+) \to \infty$& 
 $(1,y^-) \to \infty$ \\
$\gamma>3\zeta_o$, $B_1<0$  &
m=1 & 
open & 
$(1,y^+) \to (0,0)$ &
 $(1,y^-) \to (0,0)$\\
  & 
&
flat  & 
$(1,y^+) \to \infty$& 
  \\
  & 
&
  & 
$(1,y^+) \to (1,y^-)$ &
  \\
  & 
&
closed  & 
$(1,y^+) \to \infty$ &
 $(1,y^-) \to \infty$ \\
   &
m=1/2   & 
open & 
$(1,y^+) \to (\bar x, \bar y)$& 
$(1,y^+) \to (0,0)$, $(1,y^-)\to  (\bar x, \bar y)$ \\
  & 
&
flat  & 
$(1,y^+) \to \infty$ &
  \\
  &
&
 & 
$(1,y^+) \to (1,y^-)$ &  
   \\
  & 
&
closed  & 
$(1,y^+) \to \infty$& 
 $(1,y^-) \to \infty$ \\
\br
 \end{tabular}
  \begin{indented}
\item [] $^{\rm a}$ {These are  exceptional trajectories and consequently do not represent typical or generic behaviour.}
  \item[] $^{\rm b}$ { In the case $m=1/2$.} 
  \label{table II}
\end{indented}
\end{table}

Typically, if $B_1<0$, then the open models evolve from the big-bang
visco-elastic singularity at $(1,y^+)$ and evolve to the Milne model at $(0,0)$ [if $m=1$] or the non-vacuum open model at $(\bar x, \bar y)$ [if $m=1/2$].
If $B_1<0$, then the closed models evolve from the big-bang visco-elastic singularity at $(1,y^+)$ to  points at infinity.  These particular points at infinity correspond to the points where $\theta=0$ (point of maximum expansion) and the various dimensionless variables breakdown.

Typical behaviour of  models with $B_1>0$ depends upon the sign of $\gamma-3\zeta_o$.  If $\gamma<3\zeta_o$, then all trajectories for the open models will violate the WEC and if $\gamma>3\zeta_o$, then the open models evolve from the big-bang visco-elastic singularity at $(1,y^+)$, become open models and then evolve towards an inflationary  flat FRW model at the point $(1,y^-)$.  A visco-elastic singularity is a singularity in which a signifigant portion of the initial total energy density is viscous elastic energy, that is $\Pi\gg 1$.  Concerning the closed models when $B_1>0$, if $\gamma<3\zeta_o$ then models evolve from the big-bang visco-elastic singularity at $(1,y^+)$  to  points at infinity.  However, if $\gamma>3\zeta_o$, then the closed models again  evolve from the big-bang visco-elastic singularity at $(1,y^+)$ but now have two different typical behaviours.  There is a class of models which approach points at infinity and do not inflate and there is a class of models which evolve towards the inflationary flat FRW model at the point $(1,y^-)$.

The flat FRW models consist of a subset of measure zero of the total state space ${\Bbb R}^5$.  However, the flat models are of a special interest
in that the flat models represent the past asymptotic behaviour of both the open and closed models.
If $B_1<0$, then the flat models evolve from the visco-elastic singularity at $(1,y^+)$ to points at infinity or to the flat model located at $(1,y^-)$.  
If $B_1>0$, and $\gamma<3\zeta_o$, then the flat models evolve from the big-bang visco-elastic singularity at $(1,y^+)$  to points at infinity.
And if $B_1>0$ and $\gamma>3\zeta_o$, then models evolve from the visco-elastic singularity at $(1,y^+)$ to points at infinity (non-inflationary) or to the inflationary model at the point  $(1,y-)$.

Note, that if the WEC is dropped (i.e., let $\gamma<3\zeta_o$) then a class of very interesting  models occurs.  There will exist models that will evolve from the visco-elastic singularity at $(1,y^+)$ with $\theta>0$ and $\dot\theta<0$ start inflating at some point $t_i$ and at a finite time after $t_i$ will start expanding at increasing rates, that is $\dot\theta>0$, and will eventually evolve towards the point $(1,y^-)$. (This is the special case mentioned in the previous subsection.)  What this means in terms of the open and flat
 models is that they
will expand with decreasing rates of expansion, start to inflate, and then continue to expand
with increasing rates of expansion.  For the closed models, the models will expand with
decreasing rates of expansion, start to inflate, and then continue to expand with increasing rates of expansion, these models will not recollapse.


\section{Conclusion\label{V}}

The only models that can possibly satisfy the WEC and inflate are those models with $\gamma>3\zeta_o$ and $B_1>0$.  Therefore, we can conclude that bulk-viscous inflation is possible in the truncated Israel-Stewart models.
However, in the models studied by Hiscock and Salmonson \cite{Hiscock91}   inflation did not occur  (note, the equations of state assumed in \cite{Hiscock91} are derived from assuming that the universe could be modelled as a Boltzmann gas), while inflation does occur in the models studied by Zakari and Jou \cite{Zakari93} who utilized different equations of state.    Furthermore, Maartens \cite{Maartens94} has also analyzed models arising from the Israel-Stewart-Hiscock theory of irreversible thermodynamics.  Maartens assumes an equation of state for the temperature $T$ of the from $T=\beta\e^{-rH_ot}$ and finds that the inflationary attractor is unstable in the case $m\geq1/2$ \cite{Maartens94}.  In our truncated model we choose dimensionless equations of state and find that inflation is sometimes possible.  The question of which equations of state are most appropriate remains unanswered, and clearly the possibility of inflation depends critically upon the equations of state utilized \cite{Zakari93}.

This work improves over previous work on viscous cosmology using the non-causal and unstable first order thermodynamics of Eckart \cite{Eckart} and differs from the work of  Belinskii et al. \cite{Belinskii80} in that dimensionless equations of state are utilized.
From the previous discussion we can conclude that the visco-elastic singularity
at the point $(1,y^+)$ is a dominant feature in our truncated models.  This singular point
remains the typical past asymptotic attractor for various values of the 
parameters $m$ and $B_1$.
This agrees with the results of Belinskii et al. \cite{Belinskii80}.
The future asymptotic behaviour depends upon both the values of $m$ and $B_1$.
If $B_1 <0$, then the open models tend to the Milne model at $(0,0)$ [$m=1$] or to the open model at $(\bar x, \bar y)$ [$m=1/2$], and if $B_1>0$, then the open models
tend to the inflationary model at the point $(1,y^-)$ or are unphysical.
If $B_1 <0$, then the closed models tend to points at infinity, and if $B_1>0$, then the closed models tend to the inflationary model at the point $(1,y^-)$. The future asymptotic behaviour of the flat models is that they either tend to points at infinity  or tend to the point $(1,y^-)$, in agreement with the exact solution  given in \cite{Chimento93}.

Belinskii et al. \cite{Belinskii80} utilized
the physical variables $\theta$, $\rho$, and $\Pi$ in their analysis and assumed non-dimensionless equations of state, and they found a singularity
in which the expansion was zero but the metric coefficients were neither
infinite nor zero;  the authors passed over this observation stating that
in a more realistic theory this undesirable asymptotic behaviour would not occur.  We note that this behaviour did not occur in our analysis in which  dimensionless variables and and a set of dimensionless equations of state were utilized.

The behaviour of the Eckart models in Burd and Coley \cite{Burd94} with $9\zeta_0-(3\gamma-2)>0$  is very similar to 
the behaviour of the truncated Israel-Stewart models studied here in the case $B_1<0$.  This result also agrees with the conclusions of Zakari and Jou \cite{Zakari93}.
However, when $B_1>0$ various new possibilities can occur; for instance, there exist open and closed models that  asymptotically approach a flat FRW model both to the past
and to the future.  Interestingly enough this is also the case in which the future asymptotic endpoint is an inflationary attractor.
This type of behaviour does not occur in the Eckart theory.

While this work (and the work of Belinskii et al. \cite{Belinskii80}) 
employs a causal and stable second order relativistic theory of thermodynamics,
only the `truncated' version of the theory has been utilized \cite{Maartens94}, rather than the full theory of Israel-Stewart-Hiscock  \cite{Hiscock91,Zakari93,IsraelStewart79,Pavon82}.  However, the `truncated' equations can give rise to very different behaviour than the full equations;  indeed, Maartens \cite{Maartens94} argues that generally the `truncated' theory will lead to pathological behaviour, particularly in the behaviour of the temperature.  It is possible to reconcile the truncated and full theories, at least near to equilibrium, by using a ``generalized'' non-equilibrium temperature  and pressure defined via a modified Gibbs equation in which dissipation enters explicitly \cite{Gariel}.  Moreover, it is expected that the truncated theory studied here is applicable at least in the early universe.
We have neglected the divergence terms here as a first approximation.  If one includes the divergence terms then the system of equations describing the model requires an additional equation of state for the temperature $T$.  A reasonable assumption in this case might be to assume a dimensionless equation of state for $T$.

Notwithstanding these comments it is clear that the next step in the study of dissipative processes in the universe is to utilize the {\it full} (non-truncated) causal theory \cite{Hiscock91,IsraelStewart79,Zakari93,Pavon82}, and this will be done in subsequent work.  In particular, we shall find that the work here will be useful in understanding the full theory and is thus a necessary first step in the analysis.  In addition,  anisotropic Bianchi type V models which   include shear viscosity and heat conduction will also be investigated.


\ack

This research was funded by the Natural Sciences and Engineering Research
Council of Canada and a Killiam scholarship awarded to RVDH. The authors would like to thank Roy Maartens for helpful discussions and for making available to us recent work prior to publication.  The authors would also like to thank Des McManus for reading the manuscript.

\appendix

\section{First Integrals}\label{appendix 1}

 We will use Darboux's theorem \cite{Schomliuk} to find an algebraic first integral of the system in the case $m={r_1}=1$ by first finding a number of a algebraic invariant manifolds.  
An algebraic invariant manifold, $Q_i=0$, is a manifold such that $\dot Q_i = r_i Q_i$, where $r_i$ is a polynomial.
The following are invariant manifolds of the system:
\begin{eqnarray}
Q_1& = & x-1, \nonumber\\
Q_2& = & y-m_-x, \nonumber \\
Q_3& = & y- m_+ x,\nonumber 
\end{eqnarray}
where
\begin{equation}
m_\pm = \frac{ (b-3\gamma) \pm \sqrt{(b-3\gamma)^2+36a}}{2}.
\end{equation}
Calculating $\dot Q_i = r_i Q_i$, we find
\begin{eqnarray}
r_1& = & -[y+(3\gamma-2)x],\nonumber \\
r_2& = & -[y+(3\gamma-2)x  + (2 -b + m_-)],\nonumber \\
r_3& = & -[y+(3\gamma-2)x + (2 -b + m_+)].\nonumber 
 \end{eqnarray}
Using Darboux's Theorem, an algebraic first integral $Q$ can be found by setting $Q= Q_1^{\:\alpha_1}Q_2^{\:\alpha_2}Q_3^{\:\alpha_3}$ and then determining what values of $\alpha_i$ satisfy the equation $\dot Q=0$.
 Solving the resulting algebraic system we find the following algebraic first integral of the dynamical system, (2.7), in the case $m={r_1}=1$: 
\begin{equation}
Q=(x-1)^{\alpha_1}(y-m_-x)^{\alpha_2}(y-m_+x)^{\alpha_3} = K \ \text{ (constant})
\end{equation}
where $\alpha_1$ is a free parameter and $\alpha_2$ and $\alpha_3$ must satisfy
\begin{eqnarray}
\alpha_2&=&\left(-\frac{1}{2}-\frac{b+3\gamma-4}{\sqrt{(b-3\gamma)^2+36a}}\right)\alpha_1,\nonumber \\
\alpha_3&=&\left(-\frac{1}{2}+\frac{b+3\gamma-4}{\sqrt{(b-3\gamma)^2+36a}}\right)\alpha_1. 
\label{alpha}
\end{eqnarray}

 This first integral determines the integral curves of the phase portraits in Figures \ref{Figure 1}, \ref{Figure 2} and \ref{Figure 3}, where the value $K$ determines which integral curve(s) is being  described.  
For example, if $K=0$, the integral curves are $x=1$ and $y=m_{\pm}x$.
Also, we can see that if $b=4-3\gamma$, $\alpha_1=-2$ and $K=1$ then the integral curve describes an ellipse; however, these closed curves necessarily
pass through the points $(1,y^+)$ and $(1,y^-)$, thereby nullifying the possible existence of closed orbits.


 \section{Numerical Analysis}\label{appendix 2}

If $m=1/2$, the dynamical system (2.7)  is  only defined for $x\geq 0$, but more importantly it is not differentiable at $x=0$.
In this section we will use  numerical techniques to analyze the integral curves in the neighborhood of the singular point $(0,0)$ in the case $m=1/2$.
  The integration and plotting was done using Maple V release 3. From the qualitative analysis we find that the behaviour depends on the parameter  $B_1$.  In the first of these two plots we choose $\gamma=1$, $a=1/9$, and $b=4$, so that $B_1=-1<0$ (see  \Fref{Figure 7}).  In the second plot we choose $\gamma=1$, $a=1/9$, and $b=2$ so that $B_1=1>0$ (see \Fref{Figure 8}).
From the numerical plots we can conclude that the point $(0,0)$ has a saddle-point like nature, in agreement with preliminary remarks made in the text in section \ref{III.2}. 

\vfill\eject

\section*{References}



\begin{thebibliography}{10}

\bibitem{Coley94a}
Coley A.A.  and van~den Hoogen R.J.  1994 {\em J. Math. Phys.} {\bf 35}(8)
  4117--4144.

\bibitem{Abolghasem93}
Abolghasem G.  and Coley A.A.  1994 {\em Int. J. Theor. Phys.} {\bf 33}(3)
  695--713.

\bibitem{Coley92}
Coley A.  and Dunn K.  1992 {\em J. Math. Phys.} {\bf 33}(5) 1772--1779.

\bibitem{Burd94}
Burd A.  and Coley A.  1994 {\em Class. Quantum Grav.} {\bf 11} 83--105.

\bibitem{MacCallum73}
MacCallum M. A.~H.  {\em {C}arg\`ese {L}ectures in {P}hysics, {\rm edited by E.
  Schatzman}}.
\newblock {G}ordon and {B}reach {N}ew {Y}ork 1973.

\bibitem{Eckart}
Eckart C.  1940 {\em Phys. Rev.} {\bf 58} 919--924.

\bibitem{Belinskii76}
Belinskii V.A.  and Khalatnikov I.M.  1976 {\em Sov. Phys. JETP} {\bf 42}(2)
  205--210.

\bibitem{Hiscock91}
Hiscock W.A.  and Salmonson J.  1991 {\em Phys. Rev. D} {\bf 43}(10)
  3249--3258.

\bibitem{Israel76}
Israel Werner  1976 {\em Ann. Phys.} {\bf 100} 310--331.

\bibitem{Israel79}
Israel W.  and Stewart J.~M.  1979 {\em Proc. R. Soc. Lond. A.} {\bf 365}
  43--52.

\bibitem{IsraelStewart79}
Israel W.  and Stewart J.~M.  1979 {\em Ann. Phys.} {\bf 118} 341--372.

\bibitem{Belinskii80}
Belinskii V.A. , Nikomarov E.S. , and Khalatnikov I.M.  1979 {\em Sov. Phys.
  JETP} {\bf 50}(2) 213--221.

\bibitem{Pavon90a}
Pav\'on D. , Bafaluy J. , and Jou D.  1991 {\em Class. Quantum Grav.} {\bf 8}
  347--360.

\bibitem{Chimento93}
Chimento L.P.  and Jakubi A.S.  1993 {\em Class. Quantum Grav.} {\bf 10}
  2047--2058.

\bibitem{Zakari93}
Zakari M.  and Jou D.  1993 {\em Phys. Rev. D} {\bf 48}(4) 1597--1601.

\bibitem{Romano94}
Romano V.  and Pav{\'o}n D.  1994 {\em Phys. Rev. D} {\bf 50}(4) 2572--2580.

\bibitem{Coley94b}
Coley A.A.  and van~den Hoogen R.J.  {\em {S}elf-similar {A}symptotic
  {S}olutions of {E}instein's {E}quations}.
\newblock {N}{A}{T}{O} {A}{S}{I} 332{B}, {P}lenum {N}ew {Y}ork 1994.

\bibitem{Coley90b}
Coley A.A.  1990 {\em J. Math. Phys.} {\bf 31}(7) 1698--1703.

\bibitem{Bluman}
Bluman G.W.  and Kumei S.  {\em {S}ymmetries and {D}ifferential {E}quations}.
\newblock {S}pringer-{V}erlag {N}ew {Y}ork 1989.

\bibitem{Coley90a}
Coley A.A.  1990 {\em Gen. Rel. Grav.} {\bf 22}(1) 3--18.

\bibitem{HawkingEllis}
Hawking S.~W.  and Ellis G.F.R.  {\em {T}he large scale structure of
  space-time}.
\newblock {C}ambridge Univ. Press {C}ambridge 1973.

\bibitem{Maartens94}
Maartens Roy  1994 {\em preprint}.

\bibitem{Pavon82}
Pav{\'o}n D. , Jou D. , and Casas-V{\'a}zquez J.  1982 {\em Ann. Inst. H.
  Poincar{\'e} A} {\bf 36} 79--88.

\bibitem{Gariel}
Gariel J.  and Denmat G.~Le  1994 {\em Phys. Rev. D} {\bf 50}(4) 2560--2566.

\bibitem{Schomliuk}
Schomliuk Dana  1993 {\em {T}rans. {A}mer. {S}oc.} {\bf 338}(2) 799--841.

\end{thebibliography}

\vfill\eject 

\Figures

\Figure{  The phase portrait describes the qualitative behavior
of the FRW models with bulk viscous pressure in the case $m={r_1}=1$ and $B_1<0$.  The arrows in the figure denote increasing $\Omega$-time ($\Omega\to\infty$) or decreasing $t$-time ($t\to 0^+$).\label {Figure 1}}

\Figure{  The phase portrait describes the qualitative behavior
of the FRW models with bulk viscous pressure in the case $m={r_1}=1$ and $B_1=0$.  The arrows in the figure denote increasing $\Omega$-time ($\Omega\to\infty$) or decreasing $t$-time ($t\to 0^+$).\label {Figure 2}}

\Figure{  The phase portrait describes the qualitative behavior
of the FRW models with bulk viscous pressure in the case $m={r_1}=1$ and $B_1>0$.  The arrows in the figure denote increasing $\Omega$-time ($\Omega\to\infty$) or decreasing $t$-time ($t\to 0^+$).\label {Figure 3}}

\Figure{  The phase portrait describes the qualitative behavior
of the FRW models with bulk viscous pressure in the case $m=1/2$ and ${r_1}=1$ with $B_1<0$.  The arrows in the figure denote increasing $\Omega$-time ($\Omega\to\infty$) or decreasing $t$-time ($t\to 0^+$).\label {Figure 4}}

\Figures

\Figure{  The phase portrait describes the qualitative behavior
of the FRW models with bulk viscous pressure in the case $m=1/2$ and ${r_1}=1$ with $B_1=0$.  The arrows in the figure denote increasing $\Omega$-time ($\Omega\to\infty$) or decreasing $t$-time ($t\to 0^+$).\label {Figure 5}}

\Figure{  The phase portrait describes the qualitative behavior
of the FRW models with bulk viscous pressure in the case $m=1/2$ and ${r_1}=1$ with $B_1>0$.  The arrows in the figure denote increasing $\Omega$-time ($\Omega\to\infty$) or decreasing $t$-time ($t\to 0^+$).\label {Figure 6}}

\Figure{  The phase portrait describes the qualitative behavior
of the FRW models in the case $m=1/2$ and $B_1=-1<0$. \label {Figure 7}}

\Figure{  The phase portrait describes the qualitative behavior
of the FRW models in the case $m=1/2$ and $B_1=1>0$. \label {Figure 8}}

\newpage

\input epsf

\begin{figure}{\[\epsfysize=10cm \epsfbox{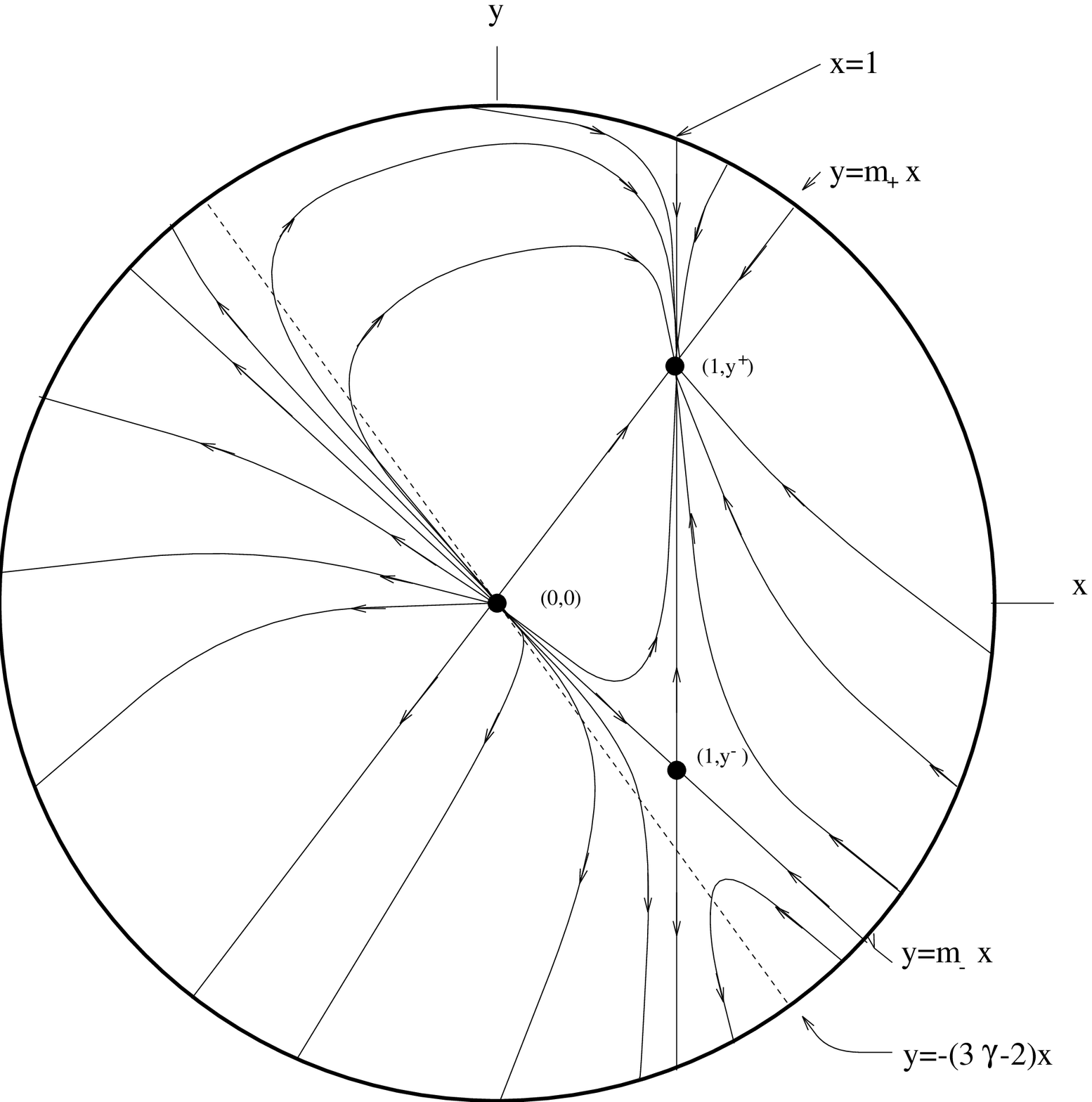}\]}
\end{figure}

\newpage
 
\begin{figure}{\[\epsfysize=10cm \epsfbox{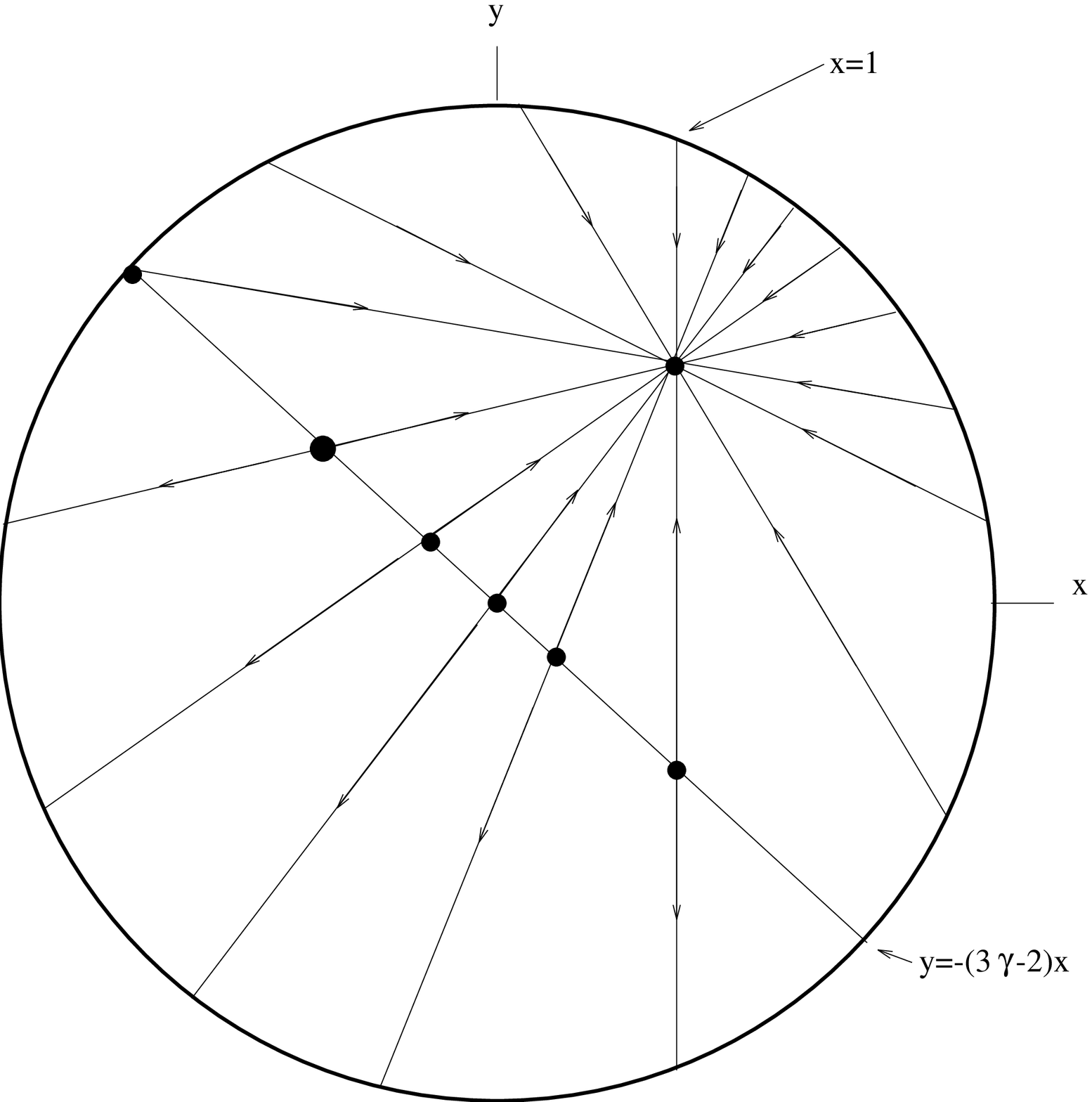}\]}
\end{figure}

\newpage

\begin{figure}{\[\epsfysize=10cm \epsfbox{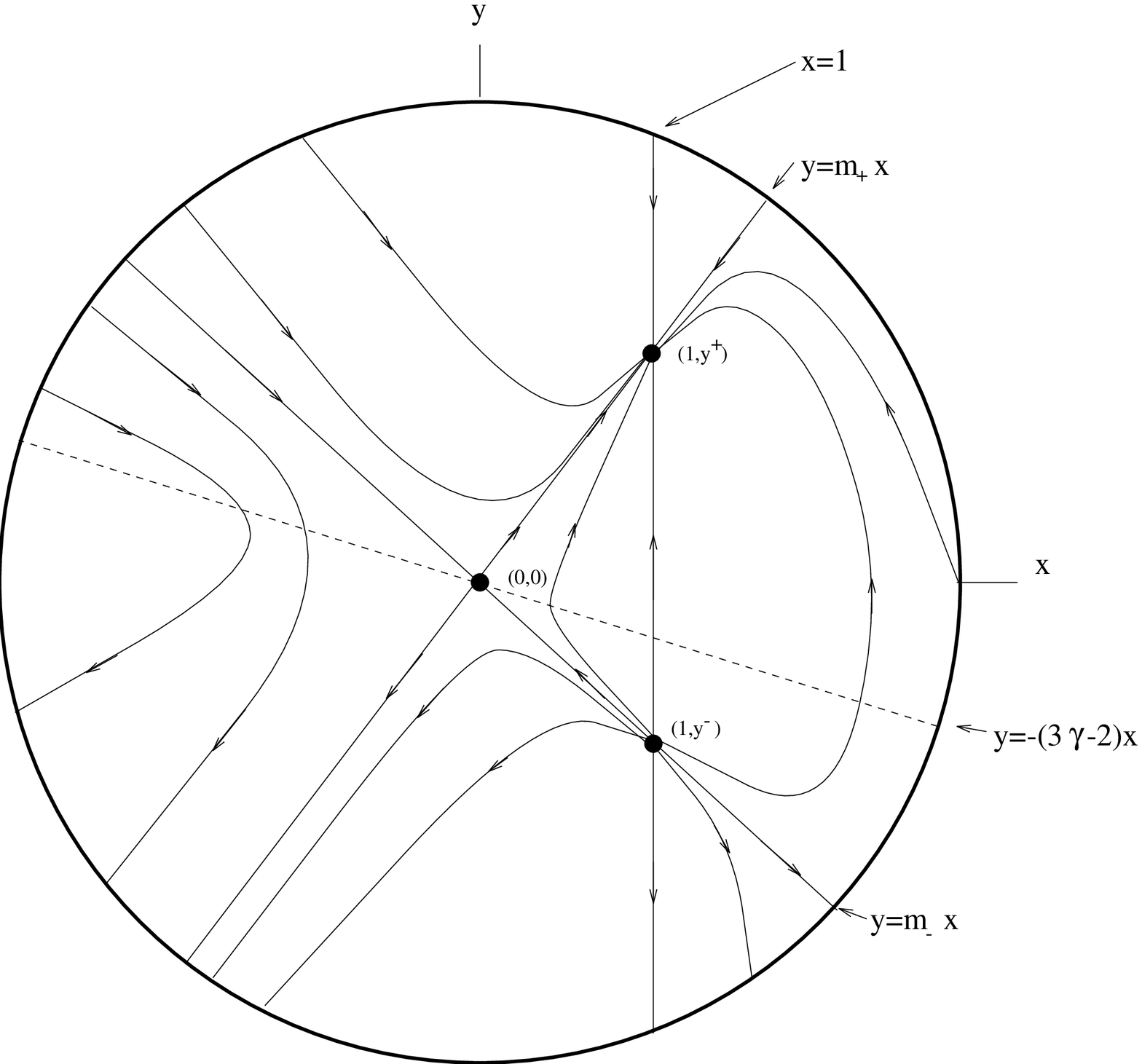}\]}
\end{figure}

\newpage

\begin{figure}{\[\epsfysize=10cm \epsfbox{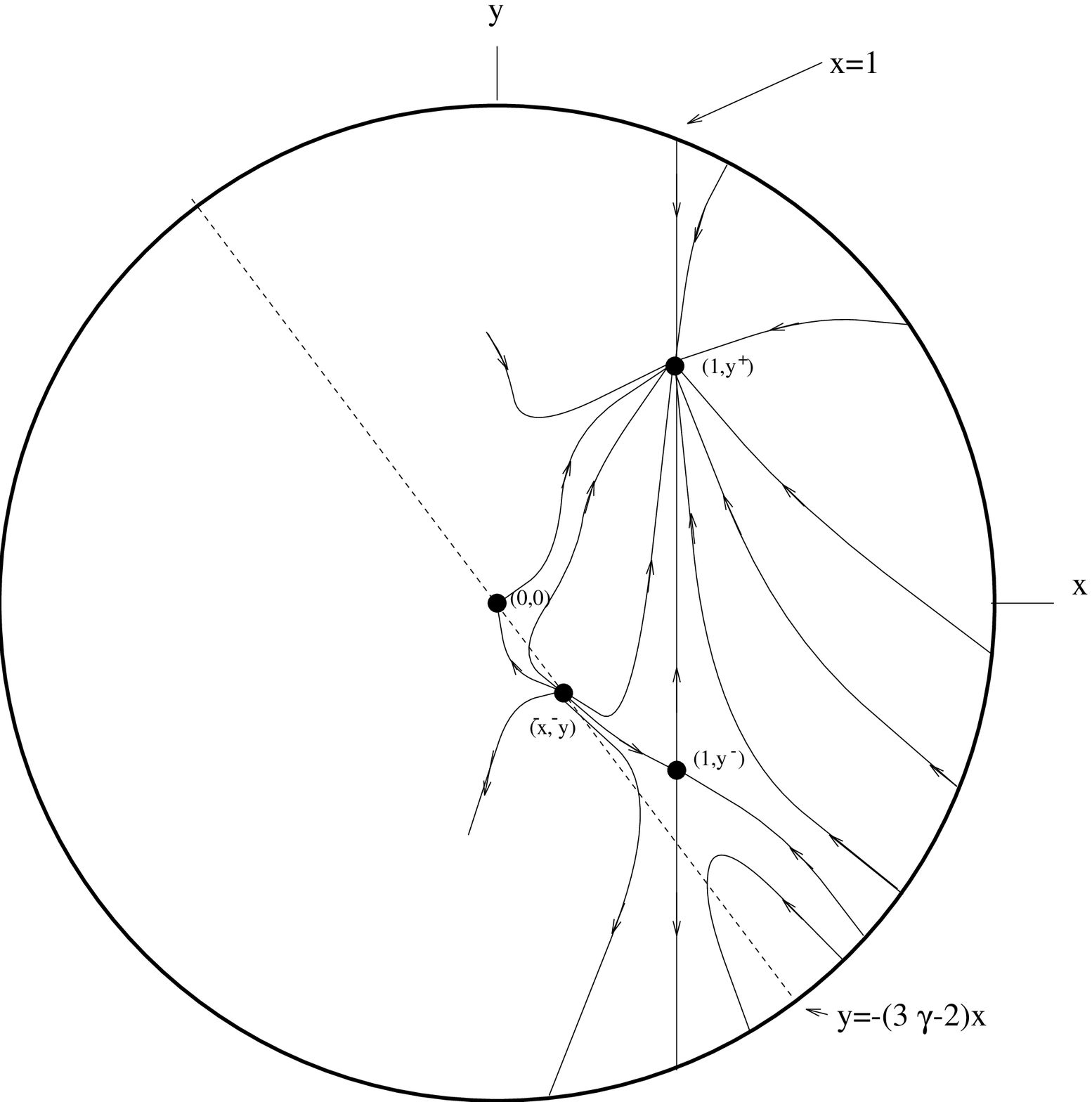}\]}
\end{figure}

\newpage

\begin{figure}{\[\epsfysize=10cm \epsfbox{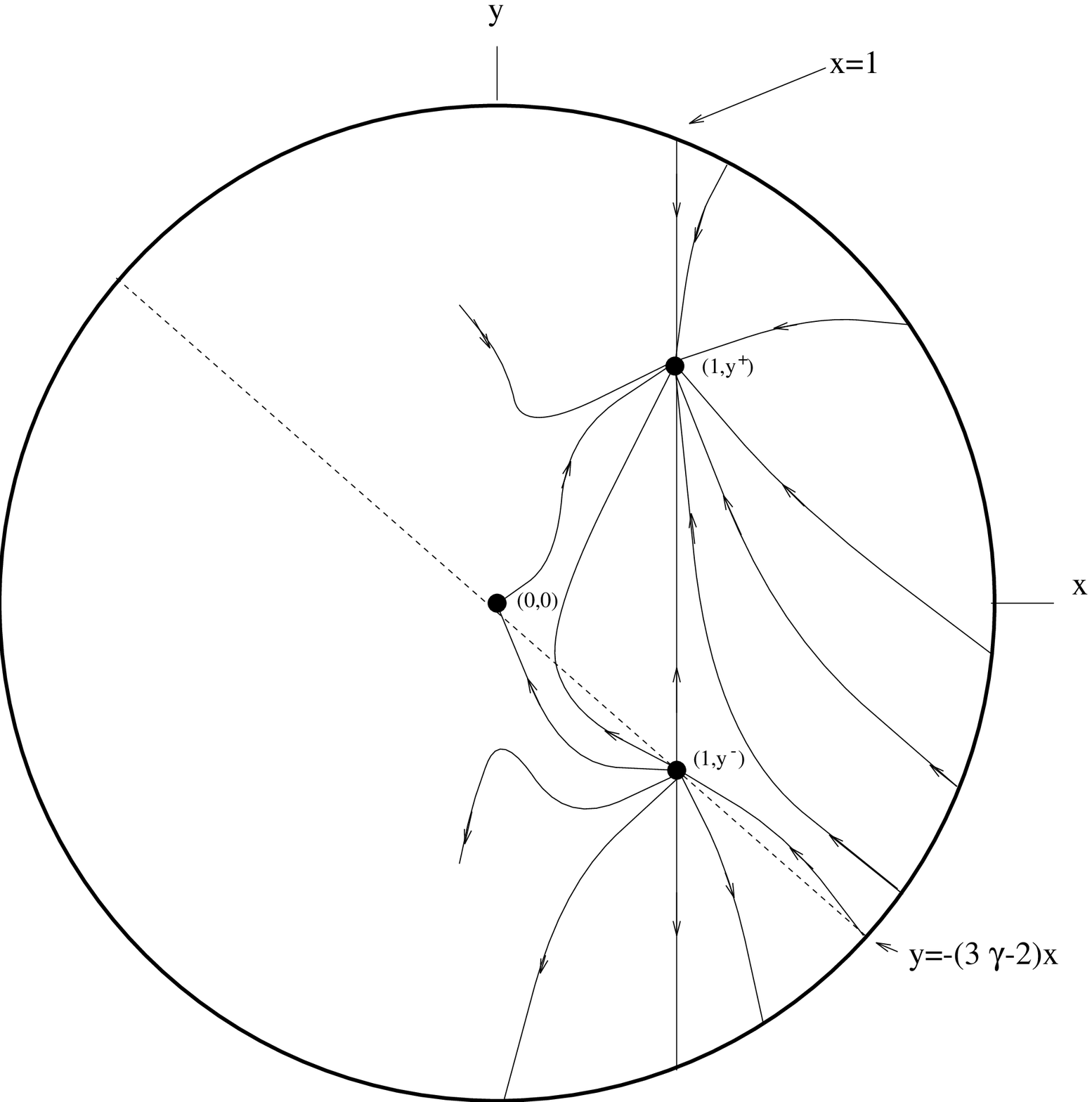}\]}
\end{figure}

\newpage

\begin{figure}{\[\epsfysize=10cm \epsfbox{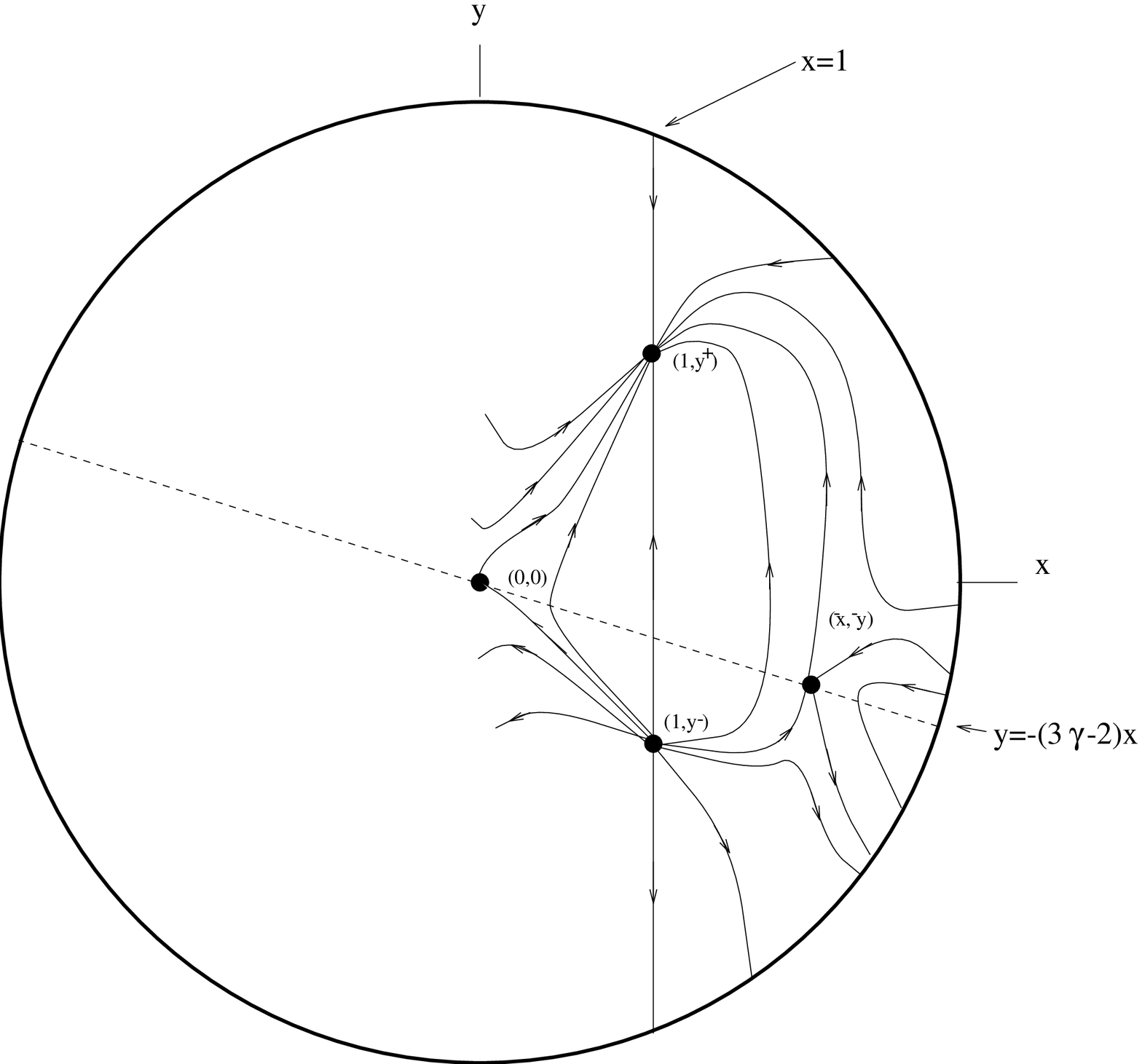}\]}
\end{figure}

\newpage

\begin{figure}{\special{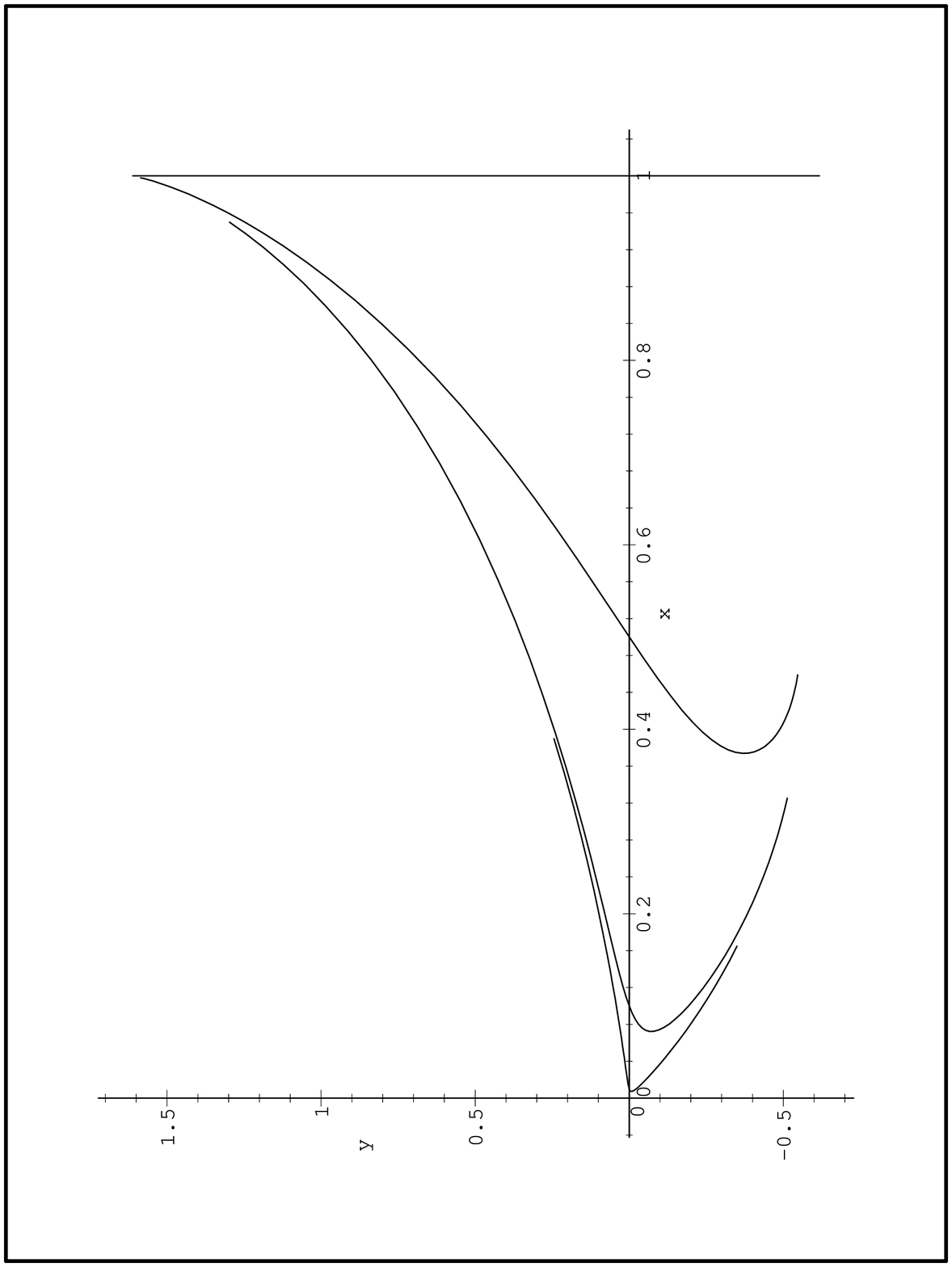 hscale=50 vscale=45 angle=270 voffset=10 hoffset=50 }}
\end{figure}

\newpage

\begin{figure}{\special{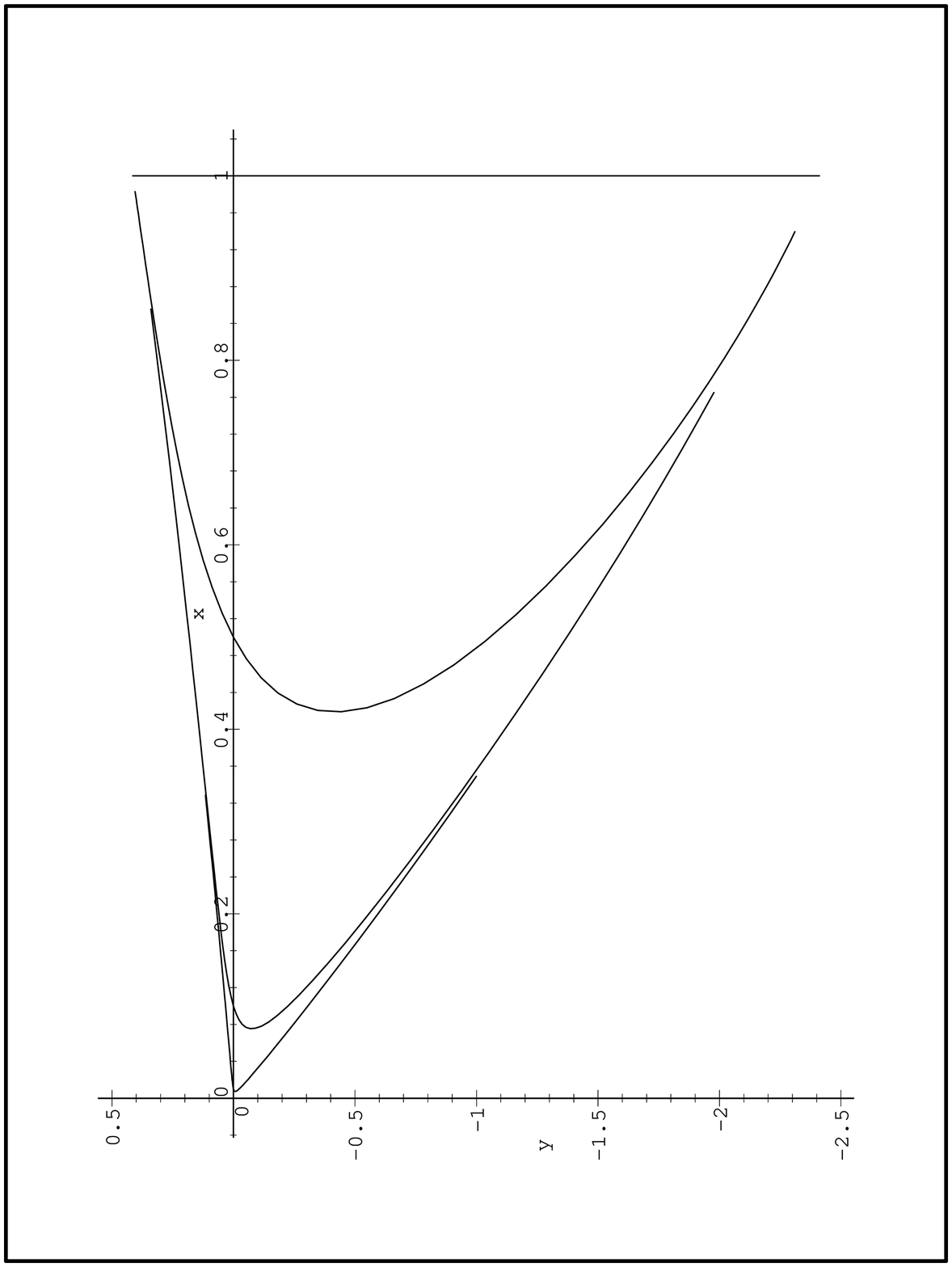 hscale=50 vscale=45 angle=270 voffset=10 hoffset=50 }}
\end{figure}

\end{document}